\documentclass[epsf,useAMS,usenatbib,usegraphicx]{mn2e}
\usepackage{amssymb}
\usepackage{amsmath}
\usepackage{times} 
\usepackage{aas_macros} 
\usepackage{booktabs}
\usepackage{lscape}
\usepackage{ulem}
\usepackage[usenames]{color}

\title[PM stars II]{Deriving the orbital properties of pulsators in binary systems through their light arrival time delays}

\author[S. J. Murphy and H. Shibahashi] 
{Simon J. Murphy$^{1,2}$ and  
Hiromoto Shibahashi$^{3}$\\
$^{1}$Sydney Institute for Astronomy, School of Physics, The University of Sydney, NSW 2006, Australia\\
$^{2}$Stellar Astrophysics Centre, Department of Physics and Astronomy, Aarhus University, 8000 Aarhus C, Denmark\\
$^{3}$Department of Astronomy, The University of Tokyo, Tokyo 113-0033, Japan \\
}

\begin{document} 

\maketitle 

\begin{abstract}%
We present the latest developments to the phase modulation method for finding binaries among pulsating stars. We demonstrate how the orbital elements of a pulsating binary star can be obtained analytically, that is, without converting time delays to radial velocities by numerical differentiation. Using the time delays directly offers greater precision, and allows the parameters of much smaller orbits to be derived. The method is applied to KIC\,9651065, KIC\,10990452, and KIC\,8264492, and a set of the orbital parameters is obtained for each system. Radial velocity curves for these stars are deduced from the orbital elements thus obtained.
\end{abstract} 


\section{Introduction} 
\label{sec:1}
Radial velocities are fundamental data of astronomy. Not only in a cosmological context, where the recessional and rotational velocities of galaxies are of interest, but also in stellar astrophysics. A time series of radial velocity (RV) data for a binary system allows the orbital parameters of that system to be calculated. However, the importance of such data, which are meticulous and time-consuming to obtain, creates a large gap between demand and supply.

In the first paper of this series \citep{murphyetal2014}, we described a method of calculating radial velocity curves using the pulsation frequencies of stars as a `clock'. For a star in a binary system, the orbital motion leads to a periodic variation in the path length travelled by light emitted from the star and arriving at Earth. Hence, if the star is pulsating, the observed phase of the pulsation varies over the orbit. We called the method `PM' for phase modulation. Equivalently, one can study orbital variations in the frequency domain, which lead to frequency modulation (FM; \citealt{FM2012}). Similar methods of using photometry to find binary stars have been developed recently by \citet{koen2014} and \citet{balona2014}, though the FM and PM methods are the first to provide a full orbital solution from photometry alone. Indeed, the application of PM to coherent pulsators will produce RV curves for hundreds of \textit{Kepler} stars without the need for ground-based spectroscopy, alleviating the bottleneck.

The crux of the PM method is the conversion of pulsational phase modulation into light arrival time delays, for several pulsation frequencies in the same star. While the phase modulation is a frequency-dependent quantity, the time delay depends on the orbital properties, only. Hence for all pulsation frequencies, the response of the time delays to the binary orbit is the same, which distinguishes this modulation from other astrophysical sources, 
such as mode interaction (see, e.g., \citealt{buchleretal1997}).

Previously, our approach used numerical differentiation of the time delays to produce a radial velocity curve, from which the final orbital solution was determined. The RV curves thus obtained were sometimes unrealistic due to scatter in the time delays. Recognising numerical differentiation as the weakness of the method, we have now developed a method of deriving the orbital properties from the time delays directly, without the need to convert time delays into RVs. It is this method that we describe in this paper. The RV curve is produced afterward, from the orbital properties, and is no longer a necessary step in the analysis.

\section{Time delay analysis: Methodology and Example 1: KIC\,9651065} 
\label{sec:2}
\begin{figure} 
\begin{center}
	\includegraphics[width=\linewidth, angle=0]{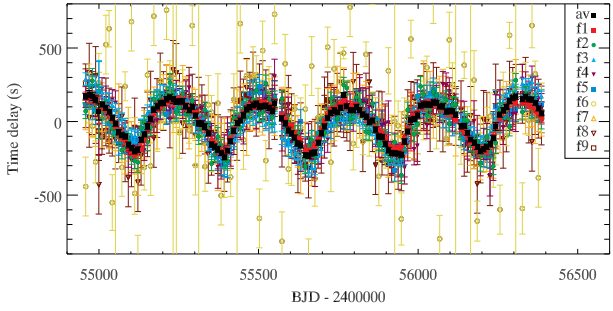}
\caption{ 
An example of TD curve (KIC\,9651065) using nine different pulsation modes, including one in the super-Nyquist frequency range \citep{sNa2013}. The weighted average is shown as solid black squares.}
\label{fig:01}
\end{center}
\end{figure}

Let us divide the light curve into short segments and measure the phase of pulsation in each segment. This provides us with `time delays' (TDs), $\tau_{\rm obs} (t_n)$, as observational data, where $t_n$ $(n=1, 2, ...)$ denotes the time series at which observations are available. Fig.\,\ref{fig:01} shows an example TD diagram (for the case of KIC\,9651065), where time delays vary periodically with the binary orbital period. The TD difference between the maximum and the minimum gives the projected size of the orbit in units of light seconds. Deviation from a sinusoid indicates that the orbit is eccentric. The TD curve is given a zero point by subtracting the mean of the time delays from each observation. The pulsating star is furthest from us when the TD curve reaches its maxima, while the star is nearest to us at the minima. 
The sharp minima and blunt maxima in Fig.\,\ref{fig:01} indicate that periapsis is at the near side of the orbit. 
The asymmetry of the TD curve, showing fast rise and slow fall, reveals that the star passes the periapsis after reaching the nearest point to us. 
In this way, TD curves provide us with information about the orbit.

Theoretically, time delay is  expressed as a function $\tau_{\rm th}(t)$ of time $t$ and the orbital elements: 
(i) the orbital period $P_{\rm orb}$, or equivalently, the orbital frequency $\nu_{\rm orb} := 1/P_{\rm orb}$, or the orbital angular frequency $\Omega := 2\upi\nu_{\rm orb}$,
(ii) the projected semi-major axis $a_1\sin i$, (iii) the eccentricity $e$, 
(iv) the angle between the nodal point and the periapsis $\varpi$,{\footnote{We have chosen to represent this angle with $\varpi$, rather than with $\omega$, because of the common use of $\omega$ to represent angular oscillation frequencies in asteroseismology.}}
and (v) the time of periapsis passage $t_{\rm p}$. Hence, these orbital elements can be determined from the observed TD as a set of parameters giving the best fitting $\tau_{\rm th}(t)$.

\subsection{Least squares method} 
\label{sec:2.1}
The best fitting parameters can be determined by searching for the minimum of the sum of square residuals
\begin{equation}
	\chi^2(\mbox{\boldmath$x$},\lambda):=
	\sum_n {{1}\over{\sigma_n^2}} \left[ \tau_{\rm th}(t_n, \mbox{\boldmath$x$}) - \tau_{\rm obs}(t_n) -\lambda\,\right]^2,
\label{eq:01}
\end{equation}
where $\sigma_n$ denotes the observational error in measurement of $\tau_{\rm obs} (t_n)$. Here the parameter dependence of $\tau_{\rm th}(t)$ is explicitly expressed with the second argument $\mbox{\boldmath$x$}$, which denotes the orbital elements as a vector, and a parameter $\lambda$ is introduced to compensate for the freedom of $\tau_{\rm obs}(t_n)=0$ (i.e. the arbitrary vertical zero-point). Hence the parameters $\mbox{\boldmath$x$}$ and $\lambda$ satisfying $\partial\chi^2/\partial\mbox{\boldmath$x$}=0$ and 
$\partial \chi^2/\partial\lambda=0$ are to be found, that is,
\begin{equation}
	\sum_n {{1}\over{\sigma_n^2}} \left[ \tau_{\rm th}(t_n, \mbox{\boldmath$x$}) - \tau_{\rm obs}(t_n) -\lambda\,\right]
	{{\partial \tau_{\rm th}(t_n, \mbox{\boldmath$x$})}\over{\partial\mbox{\boldmath$x$}}}
	=0
\label{eq:02}
\end{equation}
and 
\begin{equation}
	\lambda = \left(\sum_n {{1}\over{\sigma_n^{2}}} \right)^{-1}
 	\sum_n {{1}\over{\sigma_n^2}}[\tau_{\rm th}(t_n,\mbox{\boldmath$x$})-\tau_{\rm obs}(t_n)].
\label{eq:03}
\end{equation}

\subsection{Time delays as a function of orbital elements} 
\label{sec:2.2}
In order to solve equation (\ref{eq:02}), we have to derive the explicit dependence of $\tau_{\rm th}$ on the orbital elements. 
The readers may consult with the literature such as \cite{2001MNRAS.322..885F} (Erratum: \citealt{2009MNRAS.395.1775F}).
We derive $\tau_{\rm th}$ following \cite{FM2012} in this subsection. See also \cite{FM2}.

\begin{figure} 
\centering
\includegraphics[width=0.7\linewidth]{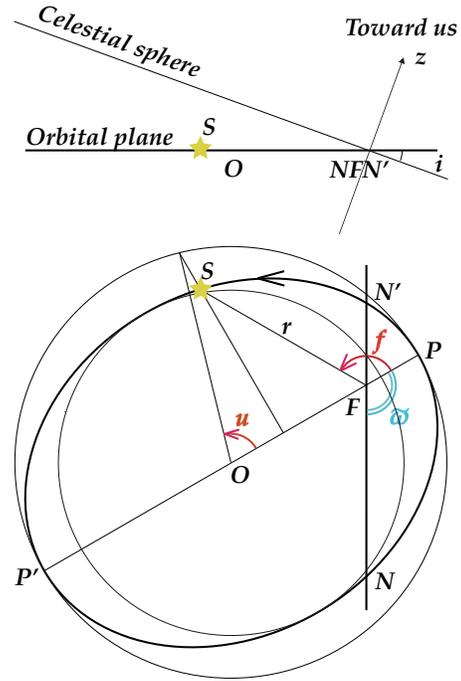} 
\caption{
{\bf Top:} Schematic side view of the orbital plane, seen from a faraway point along the intersection of the orbital plane and the celestial sphere, $NFN^\prime$, where the points $N$ and $N^\prime$ are the nodal points, respectively, and the point $F$ is the centre of gravity of the binary system; that is, a focus of the orbital ellipses. The orbital plane is inclined to the celestial sphere by the angle $i$, which ranges from $0$ to $\upi$. In the case of $0 \leq i < \upi/2$, the orbital motion is in the direction of increasing position angle of the star, while in the case of $\upi/2 < i \leq \upi$, the motion is the opposite. The $z$-axis is the line-of-sight toward us, and $z=0$ is the plane tangential to the celestial sphere.
{\bf Bottom:} Schematic top view of the orbital plane along the normal to that plane. The periapsis of the elliptical orbit is $P$. The angle measured from the nodal point $N$, where the motion of the star is directed toward us, to the periapsis {\it in the direction of the orbital motion of the star} is denoted as $\varpi$. The star is located, at this moment, at $S$ on the orbital ellipse, for which the focus is $F$. The semi-major axis is $a$ and the eccentricity is $e$. Then $\overline{{\rm OF}}$ is $ae$. The distance between the focus, $F$, and the star, $S$, is $r$. The angle $PFS$ is `the true anomaly', $f$, measured from the periapsis to the star at the moment {\it in the direction of the orbital motion of the star}. `The eccentric anomaly', $u$, also measured {\it in the direction of the orbital motion of the star}, is defined through the circumscribed circle that is concentric with the orbital ellipse. 
Figure and caption from Shibahashi et al. (2015), this Volume.
}
\label{fig:02}
\end{figure}

Let us define a plane that is tangential to the celestial sphere on which the barycentre of the binary is located, and let the $z$-axis that is perpendicular to this plane and passing through the barycentre of the binary be along the line-of-sight toward us (see Fig.\,\ref{fig:02}). The orbital plane of the binary motion is assumed to be inclined to the celestial sphere by the angle $i$, which ranges from $0$ to $\upi$. The orbital motion of the star is in the direction of {\it increasing} position angle, if $0\leq i < \upi/2$, and in the direction of {\it decreasing} position angle, if $\upi/2 < i \leq \upi$.

Let $r$ be the distance between the centre of gravity and the star when its true anomaly is $f$. The difference in the light arrival time, $\tau$, compared to the case of a signal arriving from the barycentre of the binary system is given by
\begin{equation}
	\tau = -r\sin(f+\varpi)\sin i /c
\label{eq:04}
\end{equation}
where $\varpi$ is the angle from the nodal point to the periapsis, $i$ is the inclination angle, and $c$ is the speed of the light (see Fig\,\ref{fig:02}). Note that $\tau$ is defined so that it is negative when the star is nearer to us than the barycentre.\footnote{
That convention is established as follows. When the star lies beyond the barycentre, the light arrives later than if the star were at the barycentre: it is delayed. When the star is nearer than the barycentre, the time delay is negative. A negative delay indicates an early arrival time. Since the observed luminosity, $L$, varies as
\begin{equation}
	L \sim \cos \omega(t - d/c),\nonumber
\end{equation}
where $d$ is the path length travelled by the light on its way to Earth, then the phase change, $\Delta \phi$, of the stellar oscillations is negative when the time delay is increasing. That is,
\begin{equation}
	\tau \propto - \Delta \phi. \nonumber
\end{equation}
The convention we hereby establish differs from that in PM\,I \citep[][equation 3]{murphyetal2014}, where the minus sign was not included. We therefore had to introduce a minus sign into equations (6) and (7), there, in order to follow the convention that radial velocity is positive when the object recedes from us. Hence, while the radial velocity curves in that paper have the correct orientation, the TD diagrams there are upside-down. Our convention here fixes this.
}
The distance $r$ is expressed with the help of a combination of the semi-major axis $a_1$, the eccentricity $e$, and the true anomaly $f$: \begin{equation}
	r= {{a_1(1-e^2)}\over{1+e\cos f}}.
\label{eq:05}
\end{equation}
Hence,
\begin{equation}
	\tau (t, \mbox{\boldmath$x$}) = -{{a_1\sin i}\over{c}}(1-e^2){{\sin f\cos\varpi + \cos f\sin\varpi}\over{1+e\cos f}}.
\label{eq:06}
\end{equation}
The trigonometric functions of $f$ are expressed in terms of a series expansion of trigonometric functions of the time after the star passed the periapsis with Bessel coefficients: 
\begin{equation}
	\cos f = -e + {{2(1-e^2)}\over{e}}\sum_{n=1}^\infty J_n(ne)\cos n\Omega (t-t_{\rm p}),
\label{eq:07}
\end{equation}
\begin{equation}
	\sin f = 2\sqrt{1-e^2} \sum_{n=1}^\infty J_n{'}(ne) \sin n\Omega (t-t_{\rm p}),
\label{eq:08}
\end{equation}
where $J_n(x)$ denotes the Bessel function of the first kind of integer order $n$ and $J_n{'}(x) := {\rm d}J_n(x)/{\rm d}x$. Equation (\ref{eq:06}) with the help of equations (\ref{eq:07}) and (\ref{eq:08}) gives the time delay $\tau_{\rm th}$ at time $t_n$ for a given set of $\mbox{\boldmath$x$}=(\Omega, a_1\sin i/c, e, \varpi, t_{\rm p})$.

\subsection{Simultaneous equations} 
\label{sec:2.3}
Equation (\ref{eq:02}) forms a set of simultaneous equations for the unknown $\mbox{\boldmath$x$}$ with the help of equation (\ref{eq:06}). Let us rewrite symbolically equation (\ref{eq:02}) as
\begin{equation}
	\mbox{\boldmath$y$} (\mbox{\boldmath$x$}):=\sum_n\alpha_n(\mbox{\boldmath$x$}) 
	{{\partial\tau_{\rm th}(t_n,\mbox{\boldmath$x$})}\over{\partial\mbox{\boldmath$x$}}}
	=0,
\label{eq:09}
\end{equation}
where
\begin{equation}
	\alpha_n := {{1}\over{\sigma_n^2}} 
	\left[ \tau_{\rm th}(t_n, \mbox{\boldmath$x$}) - \tau_{\rm obs}(t_n) -\lambda\,\right].
\label{eq:10}
\end{equation}
This simultaneous equation can be solved by iteration, once we have a good initial guess $\mbox{\boldmath$x$}^{(0)}$:
\begin{equation}
	\mbox{\boldmath$y$} \left(\mbox{\boldmath$x$}^{(0)}\right) 
	+ \left({{\partial \mbox{\boldmath$y$}}\over{\partial \mbox{\boldmath$x$}^{(0)}}}\right)
	\delta\mbox{\boldmath$x$}=0,
\label{eq:11}
\end{equation}
then
\begin{equation}
	\delta\mbox{\boldmath$x$}
	=
	-\left({{\partial \mbox{\boldmath$y$}}\over{\partial \mbox{\boldmath$x$}^{(0)}}}\right)^{-1}
	\mbox{\boldmath$y$} \left(\mbox{\boldmath$x$}^{(0)}\right) .	
\label{eq:12}
\end{equation}
Hence we need a means to obtain a good initial guess $\mbox{\boldmath$x$}^{(0)}$.

\begin{table} 
\centering
\caption[]{Observational constraints for KIC\,9651065.}
\begin{tabular}{lr@{\,$\pm$\,}ll}
\toprule
\multicolumn{1}{c}{Quantity} & 
\multicolumn{2}{c}{Value} & 
\multicolumn{1}{c}{Units} \\
\midrule
$\tau_{\rm max}$ & $136$ & $27$ & s \\
$\tau_{\rm min}$ & $-211$ & $42$ & s \\
$\nu_{\rm orb}$ & $0.003685$ & $0.000011$ & d$^{-1}$ \\
$A_1$ & $167.1$ & $3.06$ & s \\
$A_2$ & $35.7$ & $3.06$ & s \\
$\phi(\tau_{\rm max})$ & $0.54$ & $0.02$ & \\
$\phi(\tau_{\rm min})$ & $0.08$ & $0.02$ & \\
\bottomrule
\end{tabular}
\label{tab:01}
\end{table}

\begin{figure} 
\begin{center}
\includegraphics[width=\linewidth, angle=0]{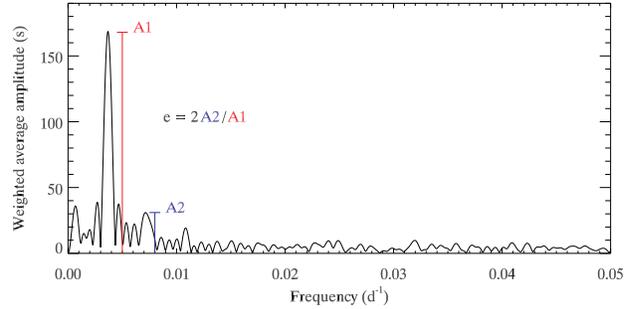} 
\caption{Fourier transform of the TD curve of KIC\,9651065 shown in Fig.\,\ref{fig:01}. 
After identification of A1 and A2 in the Fourier transform, their exact values and uncertainties are determined by a non-linear least-squares fit to the time-delay curve.} 
\label{fig:03}
\end{center}
\end{figure}

\subsection{Initial guesses} 
\label{sec:2.4}
\subsubsection{Input parameters: observational constraints} 
\label{sec:2.4.1}
An observed TD curve shows, of course, a periodic variation with the angular frequency $\Omega$. By carrying out the Fourier transform  of the observed TD curve, we determine $\Omega$ accurately. The presence of harmonics ($2\Omega$, $3\Omega$, ... ) indicates that the time delays deviate from a pure sinusoid. Hence the angular frequency $\Omega$ is fairly accurately obtained from the Fourier transform of the observed TD curve (Fig.\,\ref{fig:03}). Let $A_1$ and $A_2$ be the amplitudes in the frequency spectrum corresponding to the angular frequencies $\Omega$ and $2\Omega$, respectively. They are also accurately determined, by a simultaneous non-linear least-squares fit to the time-delay curve. By folding the observational data $\{\tau_{\rm obs}(t_n)\}$ with the period $2\upi/\Omega$, we get the time delay as a function of orbital phase, $\phi_n := \Omega (t_n-t_0)/(2\upi)$, where $t_0$ is the time of the first data point. We then know the orbital phases at which $\tau_{\rm obs}$ reaches its maximum and minimum. In the case of KIC\,9651065, shown in Fig.\,\ref{fig:01}, the frequency spectrum is shown in Fig.\,\ref{fig:03}, and the obtained quantities are summarised in Table\,\ref{tab:01}. They are used as input parameters from which initial guesses for the orbital parameters are deduced.

\subsubsection{Initial guess for $e$} 
\label{sec:2.4.2}
\begin{figure} 
\begin{center}
	\includegraphics[width=\linewidth, angle=0]{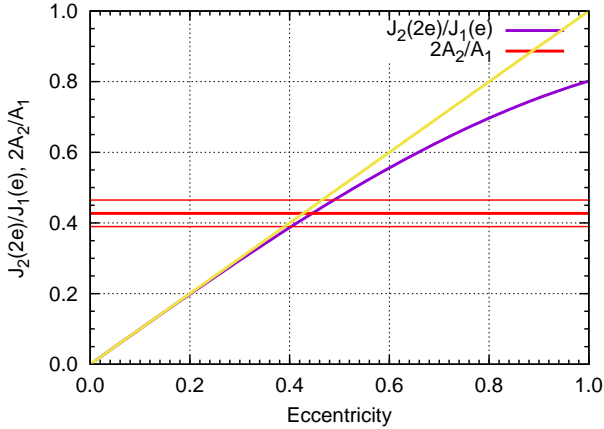} 
\caption{The amplitude ratio between the two components $A_1$ and $A_2$ of KIC\,9651065 provides us with an initial guess of $e$.
The thick horizontal red line is the measured $2A_2/A_1$, and the thin lines above and below it are the uncertainties.}
\label{fig:04}
\end{center}
\end{figure}

The amplitude ratio between the two components $A_1$ and $A_2$ provides us with an initial guess of $e$ \citep{FM2012, murphyetal2014}:
\begin{equation}
	{{J_2(2e^{(0)})}\over{J_1(e^{(0)})}} = {{2A_2}\over{A_1}} ,
\label{eq:13}
\end{equation}
where $J_1(x)$ and $J_2(x)$ denote the first kind of Bessel function, of the order of 1 and 2, respectively. 
This approximation is justified, as the $\varpi$ dependence on the amplitude ratio is weak.
In fact, this approximation is good for a wide range of $\varpi$. 
Even in the case of $e \simeq 1$, the approximation gives $e^{(0)} = 0.80$ (see Fig.\,\ref{fig:04}), from which the correct value of $e$ is recoverable.
In the case of $e \ll 1$, the LHS of the above equation is further reduced to $\sim e$ \citep{murphyetal2014}.

\subsubsection{Initial guess for $t_{\rm p}$} 
\label{sec:2.4.3}
The largest and the smallest values of $\tau_{\rm obs}$ are well defined and easily identified, as are the epochs of these extrema. Therefore the extrema and their epochs are useful for providing initial guesses for the remaining orbital elements.
 
Firstly, let us see when these extrema occur. With the help of the known laws of motion in an ellipse \citep{brouwer&clemence1961},
\begin{equation}
	r{{{\rm d}f}\over{{\rm d}t}}={{ a_1 \Omega(1+e\cos f)}\over{\sqrt{1-e^2}}}
\label{eq:14}
\end{equation}
and
\begin{equation}
	{{{\rm d}r}\over{{\rm d}t}}={{ a_1 \Omega e\sin f}\over{\sqrt{1-e^2}}},
\label{eq:15}
\end{equation}
where $\Omega$ denotes the orbital angular frequency, the time variation of $\tau$ shown in equation (\ref{eq:04}) is given by 
\begin{equation}
	{{{\rm d}\tau}\over{{\rm d}t}}=	-{{1}\over{c}} {{\Omega a_1 \sin i}\over{\sqrt{1-e^2}}} 
	\,\left[\cos (f+\varpi) + e\cos\varpi\right].
\label{eq:16}
\end{equation}
Hence, when $\tau$ reaches the extrema
\begin{equation}
	\cos (f+\varpi) = - e\cos\varpi, 
\label{eq:17}
\end{equation}
therefore
\begin{equation}
	\sin(f+\varpi) = \pm \sqrt{1-e^2\cos^2\varpi} .
\label{eq:18}
\end{equation}
Since $c\,{\rm d}\tau/{\rm d}t = v_{\rm rad}$,
the extrema of $\tau$ correspond to the epochs of $v_{\rm rad}=0$. Geometrically, in Fig.\,\ref{fig:02}, the extrema correspond to the tangential points of the ellipse to lines parallel to NN$^\prime$. Note that the nearer side corresponds to negative time delay while the farther side corresponds to positive time delay. Hereafter, the orbital elements corresponding to the extremum of the nearer side are written with a subscript `Near', and those of the farther side are distinguished with a subscript `Far'. These two points are rotationally symmetric with respect to the centre of the ellipse, O. Hence the eccentric anomalies of these two points, $u_{\rm Near}$ and $u_{\rm Far}$, are different from each other by $\upi$ radians:
\begin{equation}
	u_{\rm  Far} - u_{\rm Near} = \upi. 	
\label{eq:19}
\end{equation}
The eccentric anomaly $u$ is written with $\Omega$, $e$ and $t_{\rm p}$ as
\begin{equation}
	u=\Omega (t-t_{\rm p}) + 2\sum_{n=1}^\infty {{1}\over{n}} J_n(ne)\sin n\Omega (t-t_{\rm p}).
\label{eq:20}
\end{equation}
Since the initial guesses for $\Omega$ and $e$ are already available,
the eccentric anomaly $u$ in equation (\ref{eq:20}) is regarded as a function of $t$ with a free parameter $t_{\rm p}$. The epochs of the extrema of $\tau_{\rm obs}$, noted as $t_{\rm Near}$ and $t_{\rm Far}$, respectively, are observationally determined. Then, by substituting $t_{\rm Near}$ and $t_{\rm Far}$ into equation (\ref{eq:20}) and with a constraint given by equation (\ref{eq:19}), 
\begin{eqnarray}
\lefteqn{
	\Omega (t_{\rm Far}-t_{\rm Near})
}
\nonumber\\
\quad
\lefteqn{
	+2\sum_{n=1}^\infty {{1}\over{n}} J_n(ne)
}
\nonumber\\
\quad
\lefteqn{
	\times
	\left\{\sin n\Omega (t_{\rm Far}-t_{\rm p})-\sin n\Omega (t_{\rm Near}-t_{\rm p})\right\}
	-\upi = 0.
}
\label{eq:21}
\end{eqnarray}
This equation should be regarded as an equation with an unknown $t_{\rm p}$. To get a good initial guess for $t_{\rm p}^{(0)}$, we define
\begin{equation}
	\phi_{\rm p} := {{\Omega}\over{2\upi}} (t_{\rm p}-t_0).
\label{eq:22}
\end{equation}
and
\begin{eqnarray}
\lefteqn{
	\Psi (\phi_{\rm p}) :=
	2\upi (\phi_{\rm Far}-\phi_{\rm Near})
}
\nonumber\\
\lefteqn{
	+2\sum_{n=1}^\infty {{1}\over{n}} J_n(ne)
}
\nonumber\\
\lefteqn{
	\times
	\left\{\sin 2\upi n (\phi_{\rm Far}-\phi_{\rm p})-\sin 2\upi n( \phi_{\rm Near}-\phi_{\rm p})\right\}
	-\upi .
}	
\label{eq:23}
\end{eqnarray}
We search for zero points of $\Psi(\phi_{\rm p})$ for a given set of $(\phi_{\rm Near}, \phi_{\rm Far})$ and $e=e^{(0)}$, where $\phi_{\rm Far}$ and $\phi_{\rm Near}$ are the orbital phases corresponding to $\tau_{\rm max}$ and $\tau_{\rm min}$, respectively, that are already measured.

\begin{figure} 
\begin{center}
\includegraphics[width=\linewidth, angle=0]{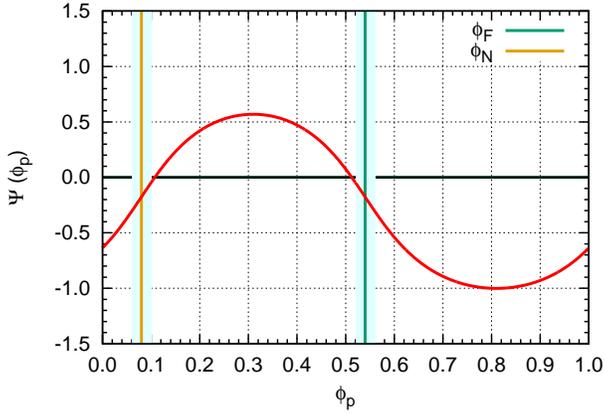} 
\caption{$\Psi(\phi_{\rm p})$ (red curve) for KIC\,9651065.
The zero crossings each give an initial estimate for $\phi_{\rm p}^{(0)}$; $\Psi(\phi_{\rm p}^{(0)})=0$. Vertical lines at `$\phi_{\rm F}$' and `$\phi_{\rm N}$' show the orbital phases of the furthest and the nearest points, corresponding to the maximum and the minimum of the time delay, respectively. The light cyan bands show the uncertainty ranges of $\phi_{\rm F}$ and $\phi_{\rm N}$.}
\label{fig:05}
\end{center}
\end{figure}

As in the case shown in Fig.\,\ref{fig:05}, there are two roots satisfying 
\begin{equation}
	\Psi \left(\phi_{\rm p}^{(0)}\right) =0,
\label{eq:24}
\end{equation}
one corresponding to the case (A) that the pulsating star in question passes the periapsis soon after the nearest point to us, and the other corresponding to the case (B) that the star passes the apoapsis just before the nearest point to us. It is expected then that the sum of $\varpi$ derived from these two solutions is $2\upi$, that is, these two solutions are explementary angles.

\begin{figure} 
\begin{center}
\includegraphics[width=\linewidth, angle=0]{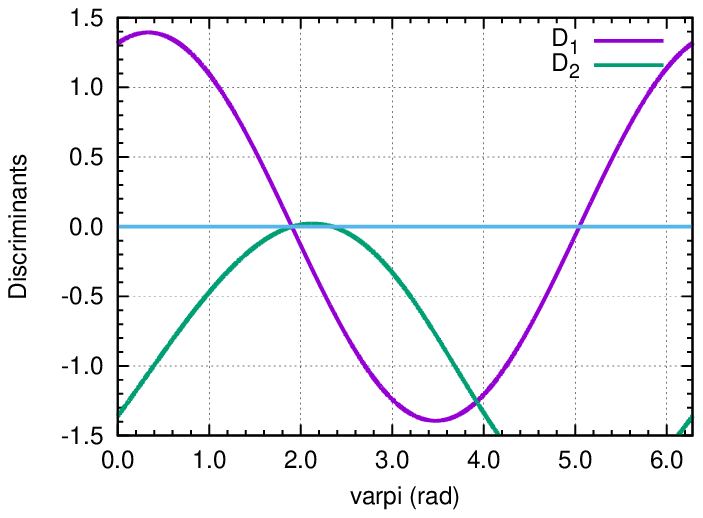}  
\includegraphics[width=\linewidth, angle=0]{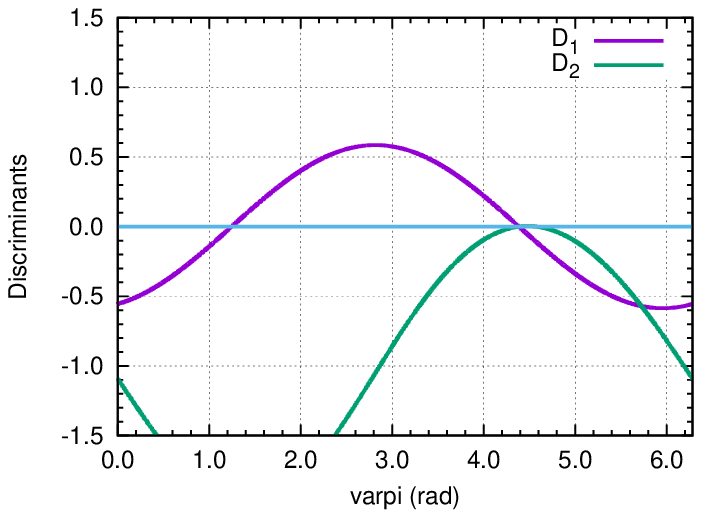}  
\caption{Discriminants from equations (\ref{eq:27}) and (\ref{eq:28}) for  $\varpi$ of KIC\,9651065. The value of $\varpi$ satisfying both of $D_1(\varpi)=0$ and $D_2(\varpi)=0$ can be the solution. 
The upper panel is the case (A) that the periapsis in the near side to us, while the lower panel indicates the case (B) that the periapsis in the far side from us. 
It is clearly seen that the angle $\varpi$ of the case (A) and that of the case (B) are explementary angles.
}
\label{fig:06}
\end{center}
\end{figure}

\begin{figure} 
\begin{center}
\includegraphics[width=0.4\linewidth, angle=0]{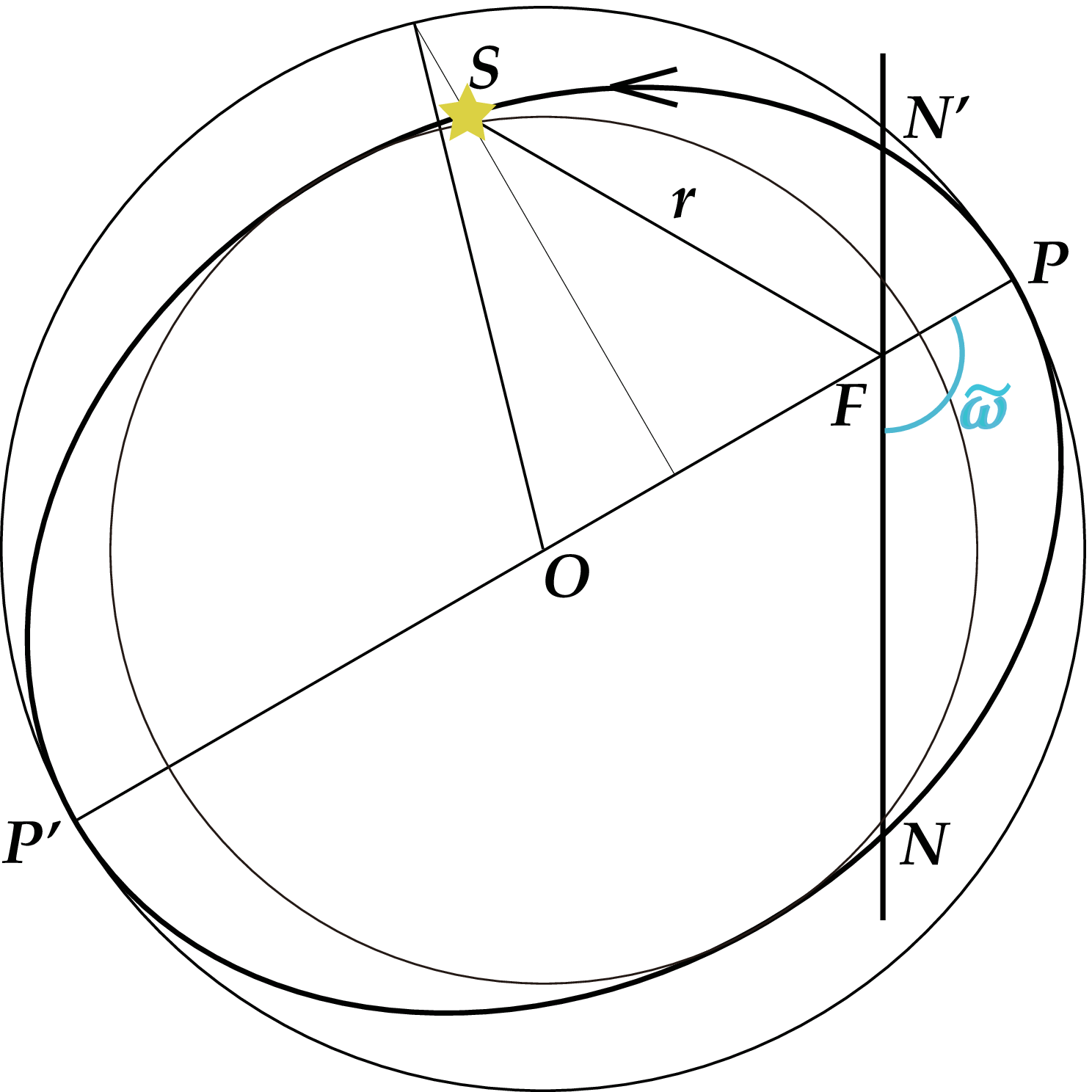}  
\hspace{1cm}
\includegraphics[width=0.4\linewidth, angle=0]{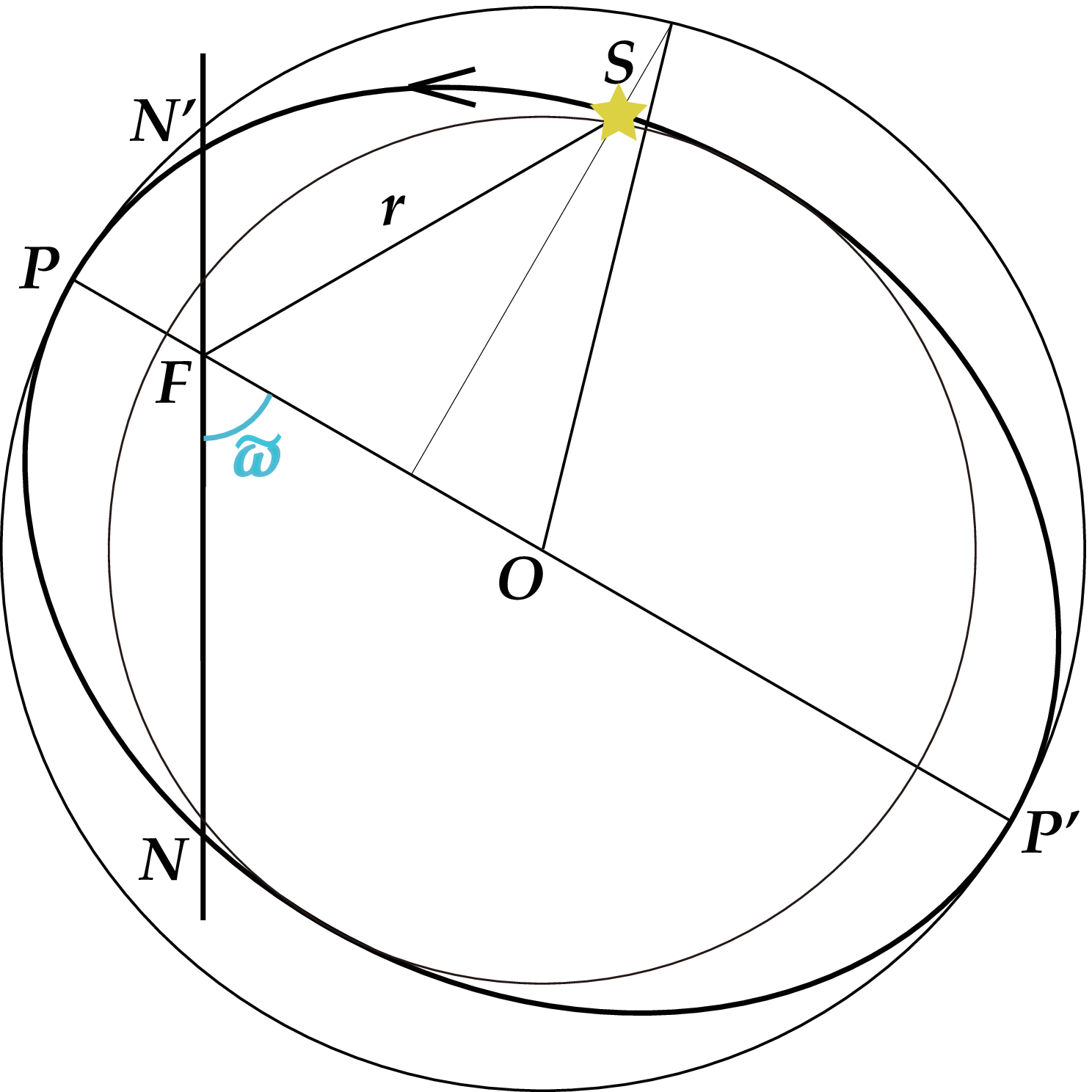} 
\caption{
Two solutions satisfying $D_1(\varpi)=D_2(\varpi)=0$.
The line of sight is assumed to be perpendicular to $NN'$ and the star is viewed from the right-hand side. 
The left panel is the case (A) that the periapsis 
in the near side to us, while the right panel indicates the case (B) that the periapsis 
in the far side from us. It is clearly seen that the solution of the case (A) and that of the case (B) are explementary angles.
}
\label{fig:07}
\end{center}
\end{figure}

\subsubsection{Initial guess for $\varpi$} 
\label{sec:2.4.4}
Once $\phi_{\rm p}$ is determined, equations (\ref{eq:07}) and (\ref{eq:08}) give the true anomaly at the nearest point, $f_{\rm Near}$;
\begin{equation}
	\cos f_{\rm Near}^{(0)} = -e + {{2(1-e^2)}\over{e}}\sum_{n=1}^\infty J_n(ne)\cos 2\upi n (\phi_{\rm Near}-\phi_{\rm p}),
\label{eq:25}
\end{equation}
\begin{equation}
	\sin f _{\rm Near}^{(0)} = 2\sqrt{1-e^2} \sum_{n=1}^\infty J_n{'}(ne) \sin 2\upi n (\phi_{\rm Near}-\phi_{\rm p}).
\label{eq:26}
\end{equation}

Since equations (\ref{eq:17}) and (\ref{eq:18}) should be satisfied at the nearest point, we define two discriminants
\begin{equation}
	D_1(\varpi) := \cos(f_{\rm Near}+\varpi)+e\cos\varpi
\label{eq:27}
\end{equation}
\begin{equation}
	D_2(\varpi) := \sin(f_{\rm Near}+\varpi) - \sqrt{1-e^2\cos^2\varpi}
\label{eq:28}
\end{equation}
and search for $\varpi^{(0)}$ satisfying both of $D_1(\varpi^{(0)})=0$ and $D_2(\varpi^{(0)})=0$ (Fig.\,\ref{fig:06}). Corresponding to the presence of two possible solutions of $\phi_{\rm p}$, there are two solutions of $\varpi$, 
which are explementary angles (see Fig.\,\ref{fig:07}). 

\begin{figure} 
\begin{center}
\includegraphics[width=\linewidth, angle=0]{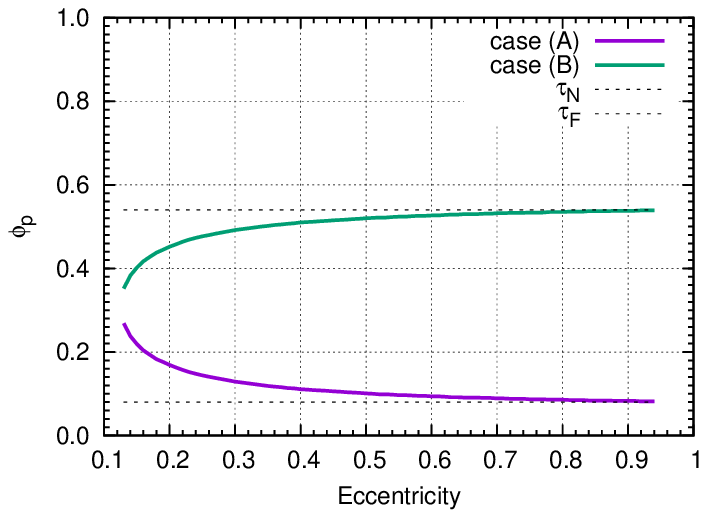} 
\includegraphics[width=\linewidth, angle=0]{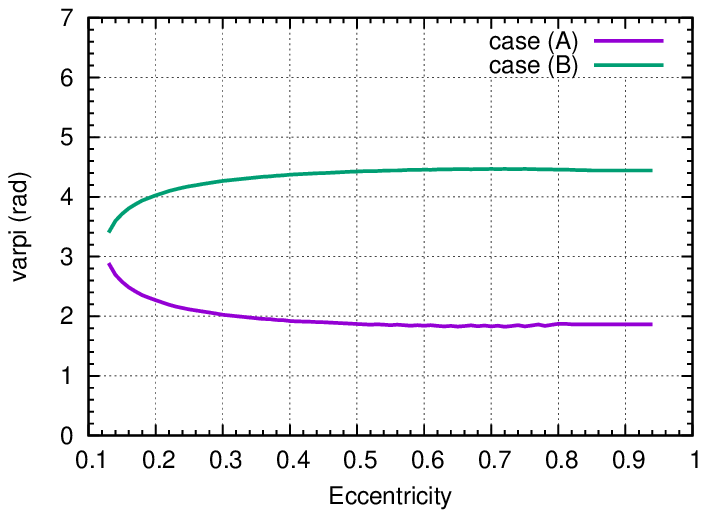} 
\caption{The dependence of $\phi_{\rm p}$ (top panel) and $\varpi$ (bottom panel) on $e$ 
for a given set of $\tau_{\rm F}$ and $\tau_{\rm N}$ of KIC\,9651065.  
In this case, for $e \lesssim 0.13$, there is no solution satisfying $\Psi(\phi_{\rm p})=0$.
It is clearly seen that $\varpi$ of the case (A) and that of the case (B) are explementary angles.
}
\label{fig:08}
\end{center}
\end{figure}

It should be noted here that both $\phi_{\rm p}$ and $\varpi$ are determined as functions of $e$. Their dependence on $e$ for a given set of $\tau_{\rm F}$ and $\tau_{\rm N}$ is shown in Fig.\,\ref{fig:08}.

\subsubsection{Initial guess for $a_1\sin i$} 
\label{sec:2.4.5}
Once $e$ and $\varpi$ are determined, 
the projected semi-major axis, $a_1\sin i$, is determined in units of light travel time, 
with the help of $\tau_{\rm max}-\tau_{\rm min}$, by
\begin{equation}
	{{a_1\sin i}\over{c}} = {{(\tau_{\rm max}-\tau_{\rm min})}\over{2}}
		\left(1-e^2\cos^2\varpi\right)^{-1/2}.
\label{eq:29}
\end{equation}
Note that the two solutions of $\varpi$ obtained above lead to an identical value of $a_1\sin i/c$.

\subsubsection{TD curve for initial guesses} 
\label{sec:2.4.6}
\begin{table} 
\centering
\caption[]{Possible solutions as initial guesses for the orbital parameters of KIC\,9651065 deduced from the observational constraints listed in Table\,\ref{tab:01}. The parameters $\phi_{\rm p}$ and $\varpi$ given in the first line of each are appropriate to be initial guesses, while  those in the second line are unsuitable.}
\begin{tabular}{lr@{\,$\pm$\,}ll}
\toprule
\multicolumn{1}{c}{Quantity} & 
\multicolumn{2}{c}{Value} & 
\multicolumn{1}{c}{Units} \\
\midrule 
$\nu_{\rm orb}$ & $0.003685$ & $0.000011$ & d$^{-1}$ \\ 
$(a_1\sin i)/c$ & $174$ & $25$ & s \\
$e$ & $0.427$ & $0.037$ & \\
\midrule
$\phi_{\rm p}$ & $0.11$ & $0.04$ &  \\
                       & \multicolumn{2}{c}{\it 0.51}  & \\
$\varpi$ & $1.90$ & $0.23$ & rad \\
                 &  \multicolumn{2}{c}{\it 4.39}  & \\
\bottomrule
\end{tabular}
\label{tab:02}
\end{table}

\begin{figure} 
\begin{center}
	\includegraphics[width=0.95\linewidth, angle=0]{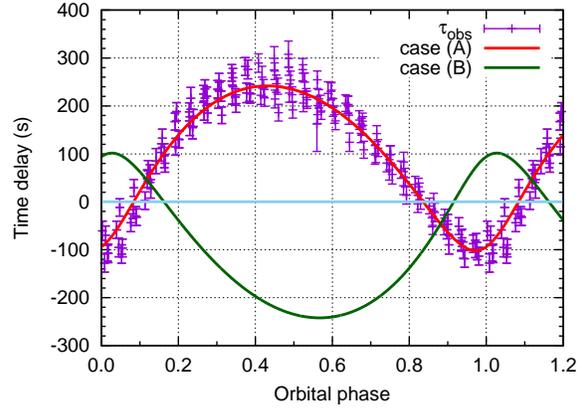}  
\caption{The TD curves for KIC\,9651065 constructed with the two sets of initial guesses for the orbital parameters. The red curve, generated with the parameters in the first line of Table\,\ref{tab:02}, matches the observed `time delay' $\tau_{\rm obs}$ (violet dots with error bars), wrapped with the orbital period. 
On the other hand, the green curve generated with the parameters in the second line of Table\,\ref{tab:02} has a larger value of  $\chi^2/ N$, 
so it is rejected. The periapsis passage $\phi_{\rm p}$ was chosen as the orbital phase of zero. The data points $\tau_{\rm obs}$ were shifted vertically by the amount $\lambda$ so that they match the red curve.
}
\label{fig:09}
\end{center}
\end{figure}

Substitution of the initial guesses thus obtained into equation (\ref{eq:06}) leads to an initial guess for the TD curve. Among the two possible pairs of solutions, one of them generates a reasonable TD curve that fits the observations, while the other generates a  TD curve that is an almost mirror image of the observed TD curve. The $\chi^2$ value easily discriminates between the two values of $\varpi$, so this can be automated. Fig.\,\ref{fig:09} demonstrates the situation, using the initial guesses for the orbital parameters tabulated in Table\,\ref{tab:02}. One of the solutions, with periapsis at the far side, fits the data well, 
while the other one having periapsis at the near side has a larger value of $\chi^2/ N$,  
so the latter is rejected. Of course, the correct solution is consistent with qualitative expectations described in Sect.\,1; the periapsis of the star is at the near side of the orbit, and the pulsating star passes the periapsis after reaching the nearest point to us, that is, $\upi/2 < \varpi < \upi$.
  
Fig.\,\ref{fig:09} demonstrates how well the TD curve computed from the initial guesses reproduces the observed TDs. The orbital phase of zero is chosen so that $\phi_{\rm p}=0$. The data points of $\tau_{\rm obs}$ were vertically shifted by the amount $\lambda$ defined by equation (\ref{eq:03}), so that they match $\tau_{\rm th}$.

\subsection{Search for the best fitting parameters} 
\label{sec:2.5}
\begin{figure} 
\begin{center}
	\includegraphics[width=\linewidth, angle=0]{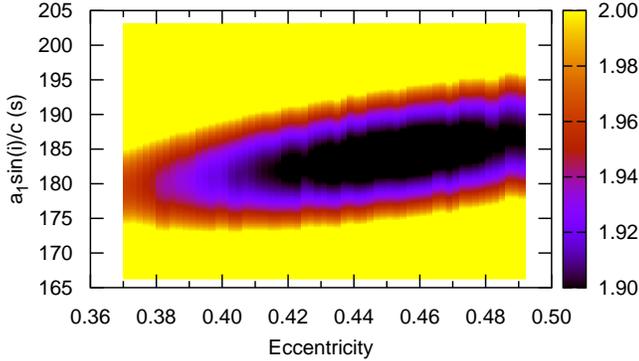} 
\caption{$\chi^2/N$ as a function of $(e, a_1\sin i/c)$ for KIC\,9651065.
}
\label{fig:10}
\end{center}
\end{figure}

\begin{figure} 
\begin{center}
	\includegraphics[width=\linewidth, angle=0]{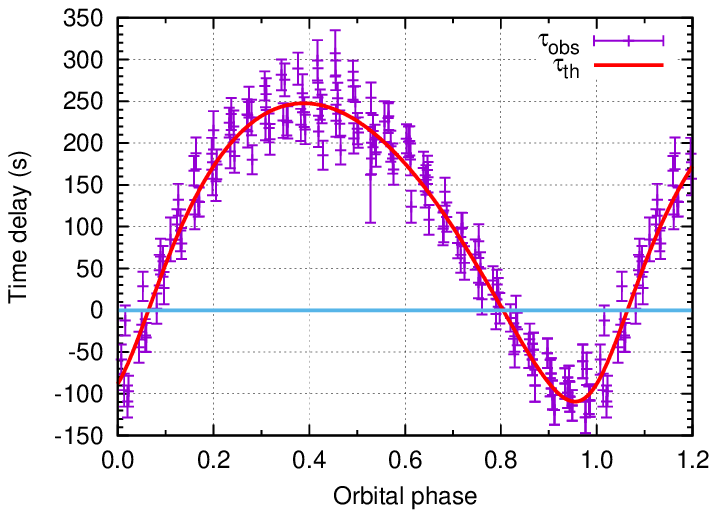}  
\caption{The best fitting TD curve for KIC\,9651065. The periapsis passage $\phi_{\rm p}$ was chosen as the orbital phase of zero.
}
\label{fig:11}
\end{center}
\end{figure}

Once a set of initial guesses for $a_1\sin i$, $e$, $\phi_{\rm p}$ and $\varpi$ are obtained, we may search for the best fitting values of these parameters that minimize $\chi^2/N$ by iteration. We regard $\Omega$ as a fixed value, because the orbital period is already well determined from the Fourier transform of the TD curve. 
The best fitting values of the orbital parameters are summarised in Table\,\ref{tab:03}, and the TD curve obtained thusly, matching best the observed time delay 
according to the $\chi^2$ minimization illustrated in Fig.\,\ref{fig:10},
is shown in Fig.\,\ref{fig:11}.

The bottom line of Table\,\ref{tab:03} lists the mass function $f(m_1,m_2,\sin i)$, defined by
\begin{eqnarray}
\lefteqn{
	f(m_1,m_2,\sin i) := {{(m_2\sin i)^3}\over{(m_1+m_2)^2}}
}
\nonumber\\
\lefteqn{
\quad\quad\quad\quad\quad\quad
	= {{(2\upi)^2c^3}\over{G}} \nu_{\rm orb}^2 \left( {{a_1\sin i}\over{c}}\right)^3,
}
\label{eq:30}
\end{eqnarray}
where $m_1$ and $m_2$ denote the masses of the primary (the pulsating star in the present case) and the secondary stars, respectively, and $G$ is the gravitational constant. The value of $f(m_1, m_2, \sin i)$ gives a minimum secondary mass of $m_2 = 0.82^{+0.04}_{-0.06},{\rm M}_{\odot}$ based on an assumption of $m_1 = 1.7\,{\rm M}_{\odot}$, so the secondary is probably a main sequence G star.

\begin{table} 
\centering
\caption[]{The best fitting orbital parameters of KIC\,9651065 deduced from the observational constraints listed in Table\,\ref{tab:01}.}
\begin{tabular}{lr@{\,$\pm$\,}ll}
\toprule
\multicolumn{1}{c}{Quantity} & 
\multicolumn{2}{c}{Value} & 
\multicolumn{1}{c}{Units} \\
\midrule 
$\nu_{\rm orb}$ & $0.003684$ & $0.000011$ & d$^{-1}$ \\ 
$(a_1\sin i)/c$ & $183.2$ & $5.0$ & s \\
$e$ & $0.44$ & $0.02$ & \\
$\phi_{\rm p}$ & $0.14$ & $0.02$ &  \\
$\varpi$ & $2.11$ & $0.05$ & rad \\
$f(m_1,m_2,\sin i)$ & $0.0896$ & $0.0074$ & M$_\odot$ \\
\bottomrule
\end{tabular}
\label{tab:03}
\end{table}

\subsection{Uncertainties} 
\label{sec:2.6}
The uncertainties on the final orbital parameters are the result of propagation of the observational uncertainties, which are obtained as follows. The uncertainty on the orbital frequency is obtained from a non-linear least-squares fit to the TD curve before phase-folding.

As for $a_1\sin i/c$ and $e$, first, we take a 2-d slice cut of $\chi^2$ in an $(e, a_1\sin i/c)$-plane 
(Fig.\,\ref{fig:10}).
Then by taking a 1-d cut of the plane at the values corresponding to the best fitting value of $a_1\sin i/c$, we get a histogram of $\chi^2$ along that line.
Since the distribution about that line is approximately Gaussian, the FWHM of that Gaussian gives the uncertainty.
The uncertainty thus evaluated on $a_1\sin i/c$ 
for KIC\,9651065
is $\sim 5$\,s,
and that on $e$ is 0.02.
The uncertainties on the other parameters are evaluated in the same manner, and they are listed in Table\,\ref{tab:03}.

\subsection{Radial velocity} 
\label{sec:2.7}
\begin{figure} 
\begin{center}
	\includegraphics[width=\linewidth, angle=0]{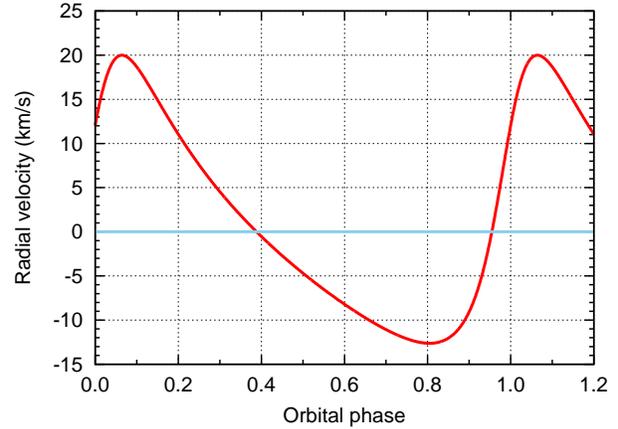} 
\caption{Radial velocity of KIC\,9651065. The periapsis passage $\phi_{\rm p}$ was chosen as the orbital phase of zero.}
\label{fig:12}
\end{center}
\end{figure}

Since $v_{\rm rad} = c\,{{\rm d}\tau}/{{\rm d}t}$, 
once the orbital parameters are deduced, it is straightforward to obtain the radial velocity:
\begin{equation}
	v_{\rm rad} =	- {{\Omega a_1 \sin i}\over{\sqrt{1-e^2}}} 
	\,\left[\cos (f+\varpi) + e\cos\varpi\right].
\label{eq:31}
\end{equation}
Fig.\,\ref{fig:12} shows the radial velocity curve thus obtained for KIC\,9651065.

In \cite{murphyetal2014}, we wrote ``{\it RV curves derived with the PM method could be used as input for codes that model eccentric binaries, such as PHOEBE. Given that such codes aim to infer the geometry of the orbit, modelling the time delays themselves might be preferred over the RV curve, since the former give the binary geometry directly and more precisely.}'' Our hopes were realized in the present work. The radial velocity curve is now provided only as a visualisation, ---it is not required for the derivation of the orbital parameters.

\section{Example 2: KIC\,10990452} 
\label{sec:3}
\begin{figure} 
\begin{center}
	\includegraphics[width=\linewidth]{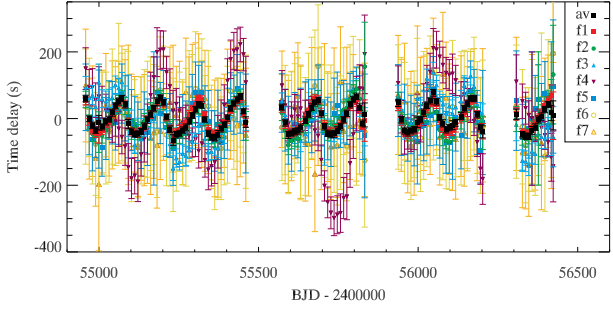} 
	\caption{
TD curve of KIC\,10990452 obtained from seven different pulsation modes. The weighted average is shown as solid black squares.} 
\label{fig:13}
\end{center}
\end{figure}

Our method is also applicable to pulsators in binary systems with short time delays.
In this section, we demonstrate KIC\,10990452, for which the range of variation in time delay is about 1/4 of the case of KIC\,9651065. Fig.\,\ref{fig:13} shows the TD curve for KIC\,10990452. Deviation from a sinusoid indicates that the orbit is eccentric as in the case of Example 1: KIC\,9651065. However, contrary to the case of KIC\,9651605, its maxima are sharp and the minima are rounded. These facts indicate that periapsis is at the far side of the orbit. Fast fall and slow rise reveal that the star passes the periapsis after reaching the farthest point from us.
Table\,\ref{tab:04} summarises the observational constraints for KIC\,10990452.

\begin{table} 
\centering
\caption[]{Observational constraints for KIC\,10990452.} 
\begin{tabular}{lr@{\,$\pm$\,}ll}
\toprule
\multicolumn{1}{c}{Quantity} & 
\multicolumn{2}{c}{Value} & 
\multicolumn{1}{c}{Units} \\
\midrule
$\tau_{\rm max}$ & $63.2$ & $12.6$  & s \\
$\tau_{\rm min}$ & $-45.4$ & $10.0$ & s \\
$\nu_{\rm orb}$ & $0.0081855$ & $0.0000142$ & d$^{-1}$ \\
$A_1$ & $49.22$ & $1.95$ & s \\
$A_2$ & $13.14$ & $1.32$ & s \\
$\phi(\tau_{\rm max})$ & $0.77$ & $0.02$ & \\
$\phi(\tau_{\rm min})$ & $0.15$ & $0.02$  & \\
\bottomrule
\end{tabular}
\label{tab:04}
\end{table}

\begin{figure} 
\begin{center}
	\includegraphics[width=\linewidth, angle=0]{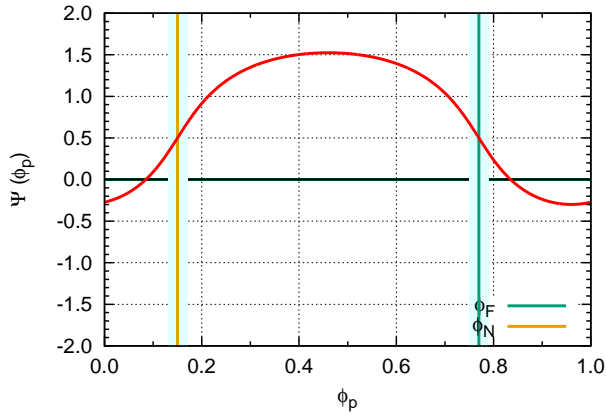} 
\caption{Same as Fig.\,\ref{fig:05}, but for KIC\,10990452.}
\label{fig:14}
\end{center}
\end{figure}

\begin{figure} 
\begin{center}
	\includegraphics[width=\linewidth, angle=0]{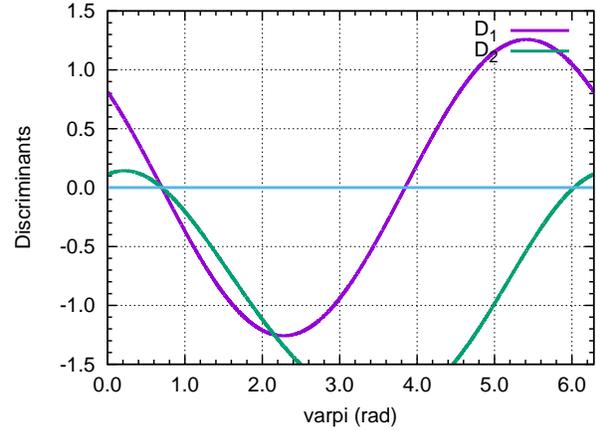} 
	\includegraphics[width=\linewidth, angle=0]{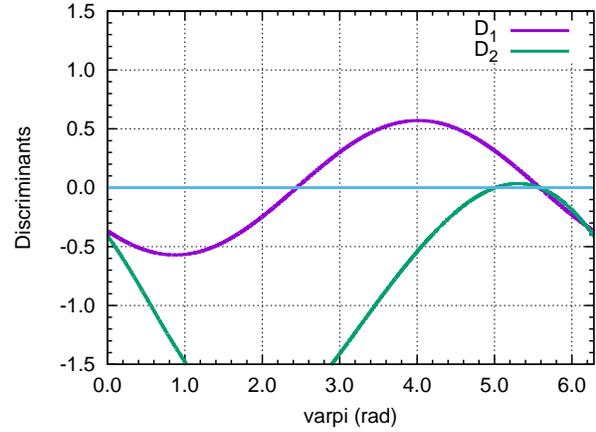} 
\caption{Same as Fig.\,\ref{fig:06}, but for KIC\,10990452.}
\label{fig:15}
\end{center}
\end{figure}

Substitution of these parameters into equation (\ref{eq:23}) leads to two roots of $\Psi(\phi_{\rm p})=0$, as shown in Fig.\,\ref{fig:14}, and each solution has $\varpi$ satisfying $D_1(\varpi)=D_2(\varpi)=0$, as demonstrated in Fig.\,\ref{fig:15}, whose sum is $2\upi$. Initial guesses for the eccentricity, $e$, and the projected semi-major axis, $a_1\sin i$, are calculated using equations (\ref{eq:13}) and (\ref{eq:29}).

\begin{figure} 
\begin{center}
	\includegraphics[width=\linewidth]{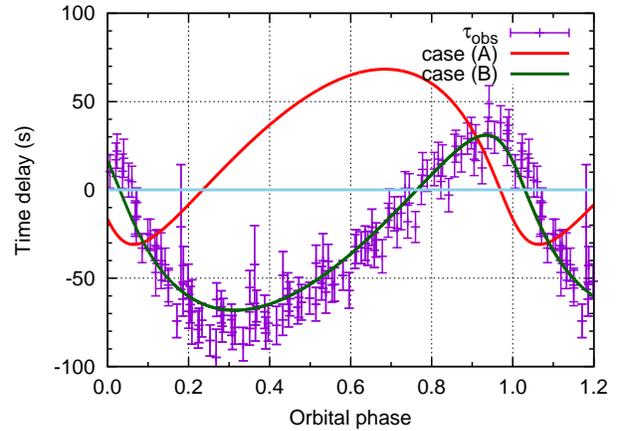}
\caption{Same as Fig.\,\ref{fig:09} but for KIC\,10990452. The green curve, generated with one of the solutions of parameters giving the smaller value of $\chi^2/N$, where $N$ denotes the number of data points, fits the data well. 
On the other hand, the red curve generated with the other set of parameters has a larger value, 
so it is rejected. The periapsis passage $\phi_{\rm p}$ was chosen as the orbital phase of zero.}
\label{fig:16}
\end{center}
\end{figure}

Substitution of the initial guesses thus obtained into equation (\ref{eq:06}) leads to an initial guess for the TD curve. Among the two possible pairs of solutions, the one giving the smaller value of $\chi^2/N$ fits the observations, as shown in Fig.\,\ref{fig:16}. The other set with the larger value of $\chi^2/N$ is rejected.

\begin{figure} 
\begin{center}
	\includegraphics[width=\linewidth, angle=0]{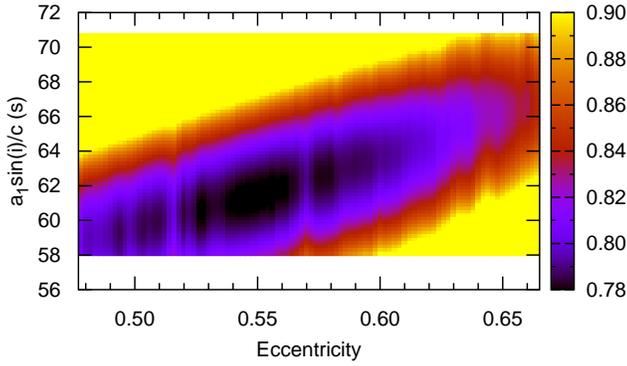} 
\caption{$\chi^2/N$ as a function of $(e, a_1\sin i/c)$ for KIC\,10990452.
}
\label{fig:17}
\end{center}
\end{figure}

\begin{figure} 
\begin{center}
	\includegraphics[width=\linewidth, angle=0]{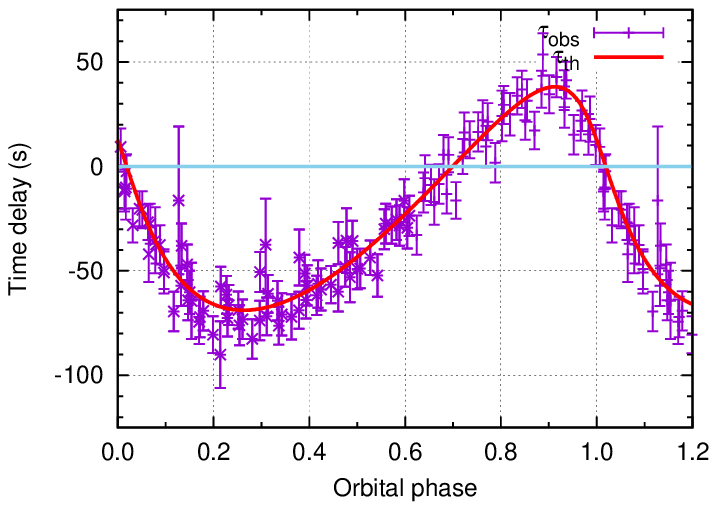} 
\caption{The best fitting TD curve for KIC\,10990452. 
The periapsis passage $\phi_{\rm p}$ was chosen as the orbital phase of zero.
}
\label{fig:18}
\end{center}
\end{figure}

\begin{figure} 
\begin{center}
	\includegraphics[width=\linewidth, angle=0]{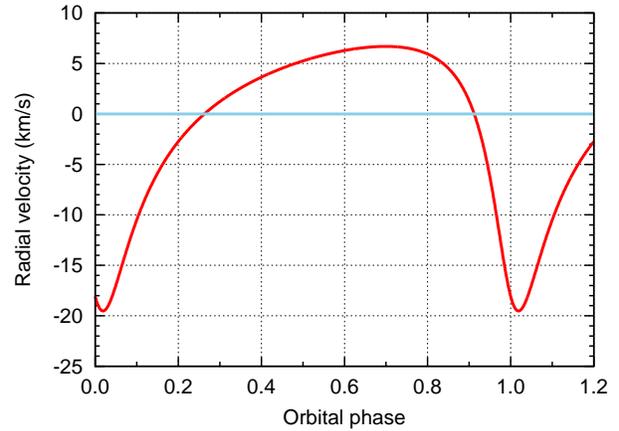}  
\caption{Radial velocity of KIC\,10990452.
The periapsis passage $\phi_{\rm p}$ was chosen as the orbital phase of zero.
}
\label{fig:19}
\end{center}
\end{figure}

The best fitting parameters are obtained by searching for the minimum of $\chi^2/N$ as a function of $(e,a_1\sin i)$. Fig.\,\ref{fig:17} shows a 2D colour map of $\chi^2/N$. The best fitting parameters are summarised in Table\,\ref{tab:05}, and the TD curve generated with these parameters is shown in Fig.\,\ref{fig:18}. Finally, the radial velocity curve is obtained as shown in Fig.\,\ref{fig:19}.

As seen in the case of KIC\,10990452, the present method is applicable without any difficulty to pulsators in binary systems showing time delay variations of several tens of seconds. Judging from the error bars in the observed time delays of \textit{Kepler} pulsators, and from the binaries we have found so far, we are confident that the present method is valid for stars showing time delay variations exceeding $\sim\pm$20\,s. While it may be possible to find binaries with even smaller time delay variations, such cases will be close to the noise level of the data and may require external confirmation. 
The noise limit is discussed further in \S\,\ref{sec:5}.

\begin{table} 
\centering
\caption[]{The best fitting orbital parameters of KIC\,10990452.}
\begin{tabular}{lr@{\,$\pm$\,}ll}
\toprule
\multicolumn{1}{c}{Quantity} & 
\multicolumn{2}{c}{Value} & 
\multicolumn{1}{c}{Units} \\
\midrule 
$\nu_{\rm orb}$ & $0.008190$ & $0.000014$ & d$^{-1}$ \\ 
$(a_1\sin i)/c$ & $61.3$ & $8.0$ & s \\
$e$ & $0.55$ & $0.03$ & \\
$\phi_{\rm p}$ & $0.89$ & $0.02$ &  \\
$\varpi$ & $5.81$ & $0.05$ & rad \\
$f(m_1,m_2,\sin i)$ & $0.01658$ & $0.00649$ & M$_\odot$ \\
\bottomrule
\end{tabular}
\label{tab:05}
\end{table}

\newpage
\section{Example 3: The more eccentric case of KIC\,8264492} 
\label{sec:4}
\begin{figure} 
\begin{center}
	\includegraphics[width=\linewidth]{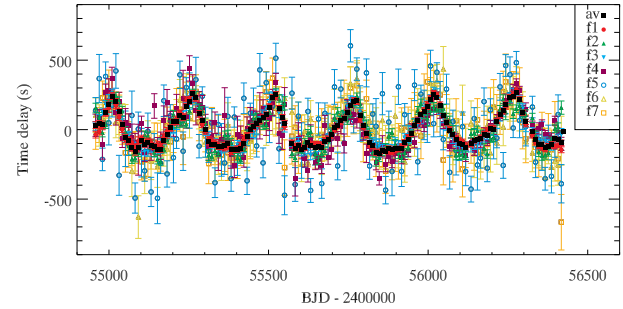}
\caption{TD curve of KIC\,8264492 obtained from seven different pulsation modes. 
The weighted average is shown as solid black squares.} 
\label{fig:20}
\end{center}
\end{figure}

\begin{figure} 
\begin{center}
	\includegraphics[width=\linewidth]{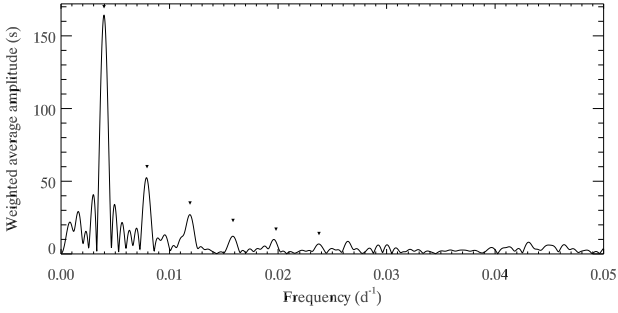} 
\caption{Fourier transform of the TD curve of KIC\,8264492 shown in Fig.\,\ref{fig:20}. Exact multiples of the orbital frequency (0.00394\,d$^{-1}$) are indicated, showing the many harmonics and implying high eccentricity.}
\label{fig:21}
\end{center}
\end{figure}

Fig.\,\ref{fig:20} shows the TD curve for another star, KIC\,8264492. 
Deviation from a sinusoid indicates that the orbit is highly eccentric, and the number of harmonics to the orbital period visible in Fig.\,\ref{fig:21}, as well as their high amplitudes, confirm this. Let us see if our method is valid for such a highly eccentric binary system. Its maxima are sharp and the minima are rounded, indicating that periapsis is at the far side of the orbit. Fast fall and slow rise reveal that the star passes the periapsis after reaching the farthest point from us.

\begin{table} 
\centering
\caption[]{Observational constraints for KIC\,8264492.} 
\begin{tabular}{lr@{\,$\pm$\,}ll}
\toprule
\multicolumn{1}{c}{Quantity} & 
\multicolumn{2}{c}{Value} & 
\multicolumn{1}{c}{Units} \\
\midrule
$\tau_{\rm max}$ & $214.6$ & $42.9$ & s \\
$\tau_{\rm min}$ &  $-132.5$ & $26.5$ & s \\
$\nu_{\rm orb}$ & $0.0039408$ & $0.0000158$  & d$^{-1}$ \\ 
$A_1$ & $159.90$ & $3.61$ & s \\
$A_2$ & $50.41$ & $3.61$ & s \\
$\phi(\tau_{\rm max})$ & $0.76$ & $0.02$ & \\
$\phi(\tau_{\rm min})$ & $0.12$ & $0.02$ & \\
\bottomrule
\end{tabular}
\label{tab:06}
\end{table}

\begin{figure} 
\begin{center}
	\includegraphics[width=\linewidth,angle=0]{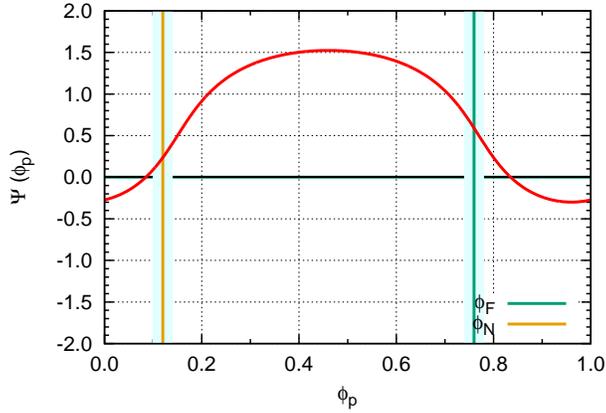} 
\caption{Same as Fig.\,\ref{fig:05}, $\Psi(\phi_{\rm p})$ (red curve), but for KIC\,8264492. 
}
\label{fig:22}
\end{center}
\end{figure}

\begin{figure} 
\begin{center}
	\includegraphics[width=\linewidth]{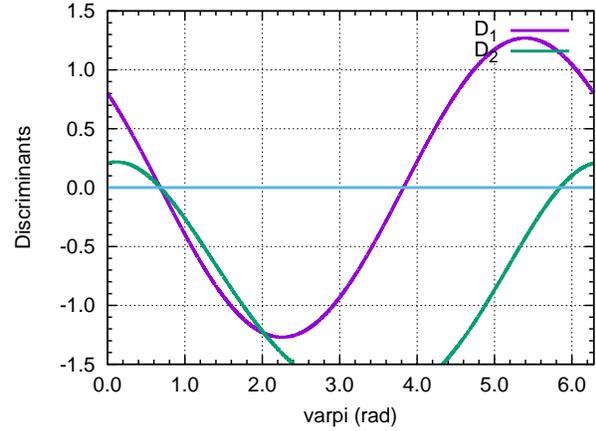} 
	\includegraphics[width=\linewidth]{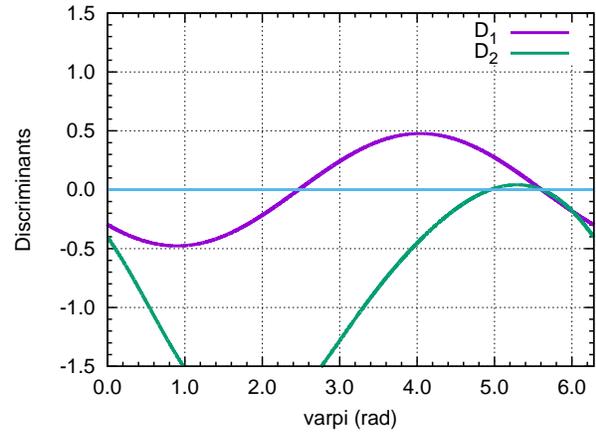} 
\caption{Same as Fig.\,\ref{fig:06} but for KIC\,8264492.
The upper panel is the case that the periapsis is in the near side to us, while the lower panel indicates the case that the periapsis 
is in the far side from us.
}
\label{fig:23}
\end{center}
\end{figure}

The orbital frequency, the amplitudes of the highest component and the second one, and the orbital phases at the maximum and the minimum of the TDs are deduced from the Fourier transform. They are summarized in Table\,\ref{tab:06}. Substitution of these parameters into equation (\ref{eq:23}) enables numerical root-finding of $\Psi(\phi_{\rm p})=0$, as shown in Fig.\,\ref{fig:22}. One root corresponds to the case (A) that the pulsating star in question passes the periapsis soon after the nearest point to us, and the other corresponds to the case (B) that the star passes the apoapsis just before the furthest point from us. Corresponding to the presence of two possible solutions of $\phi_{\rm p}$, there are two solutions of $\varpi$, which are explementary angles (see Fig.\,\ref{fig:23}).

\begin{figure} 
\begin{center}
	\includegraphics[width=\linewidth]{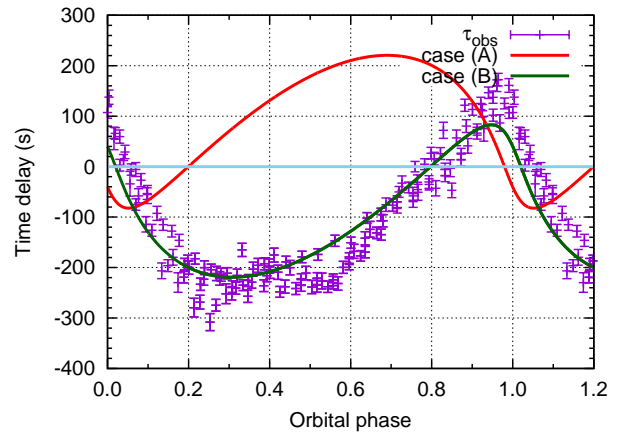} 
\caption{Same as Fig.\,\ref{fig:09} but for KIC\,8264492.
The violet curve, generated with the parameters in the first line of Table\,\ref{tab:06}, fits the data well.
On the other hand, the red curve generated 
with the other set of parameters has a larger value of $\chi^2/N$,
so it is rejected.
The periapsis passage $\phi_{\rm p}$ was chosen as the orbital phase of zero.
}
\label{fig:24}
\end{center}
\end{figure}

Substitution of the initial guesses thus obtained into equation (\ref{eq:06}) leads to an initial guess for the TD curve. As in the case of KIC\,9651605, among the two possible pairs of solutions, one of them generates a reasonable TD curve that fits the observations, 
while the other generates a TD curve that is an almost mirror image of the observed TD curve, with a larger value of $\chi^2/N$ 
(Fig.\,\ref{fig:24}). The correct solution is consistent with qualitative expectations described at the beginning of this subsection; the periapsis of the star is at the far side of the orbit, and the star passes the periapsis after reaching the farthest point from us, that is, 
$3\upi/2 < \varpi < 2\upi $.
 
Fig.\,\ref{fig:24} shows the TD curve, computed for the initial guesses, plotted with the observed TDs. The orbital phase of zero is chosen so that $\phi_{\rm p}=0$. The data points of $\tau_{\rm obs}$ were vertically shifted by the amount $\lambda$, defined by equation (\ref{eq:03}), so that they match $\tau_{\rm th}$. Unlike the earlier example of KIC\,9651065, there remain systematic residuals in the TD curve for KIC\,8264492. 

\begin{figure} 
\begin{center}
	\includegraphics[width=\linewidth, angle=0]{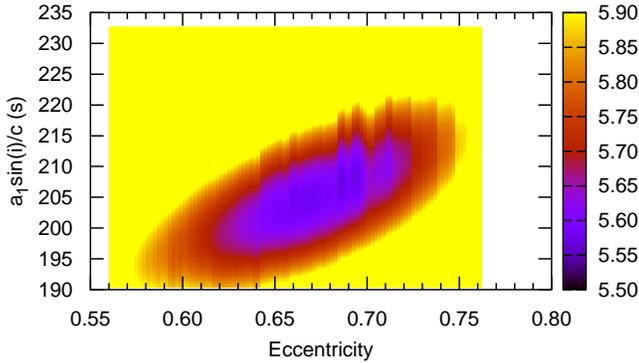} 
\caption{$\chi^2/N$ as a function of $(e, a_1\sin i/c)$ for KIC\,8264492.
}
\label{fig:25}
\end{center}
\end{figure}

\begin{figure} 
\begin{center}
	\includegraphics[width=\linewidth, angle=0]{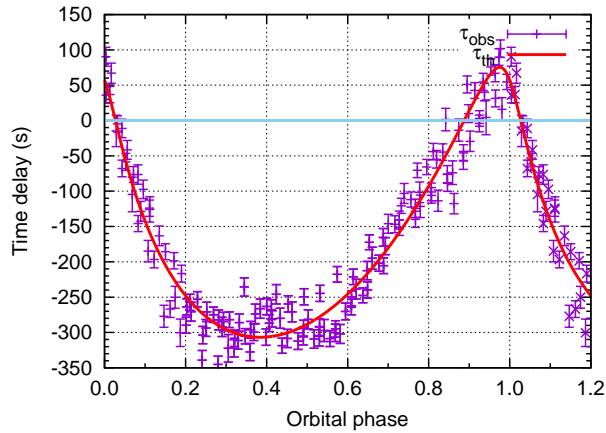} 
\caption{The best fitting TD curve for KIC\,8264492. 
The periapsis passage $\phi_{\rm p}$ was chosen as the orbital phase of zero.
}
\label{fig:26}
\end{center}
\end{figure}

\begin{figure} 
\begin{center}
	\includegraphics[width=\linewidth, angle=0]{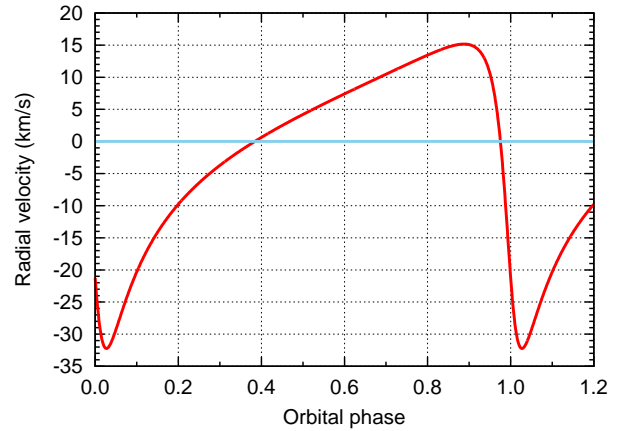} 
\caption{Same as Fig.\,\ref{fig:12} but for KIC\,8264492.
}
\label{fig:27}
\end{center}
\end{figure}

The best fitting parameters are summarised in Table\,\ref{tab:07}, and 
Fig.\,\ref{fig:26} and Fig.\,\ref{fig:27} show the TD curve and the radial velocity curve generated with these parameters, respectively.
Hence, with KIC8264492, we have demonstrated the validity and utility of the PM method, even for systems with high eccentricity.

\begin{table} 
\centering
\caption[]{The best fitting orbital parameters of KIC\,8264492 deduced from the observational constraints listed in Table\,\ref{tab:06}.}
\begin{tabular}{lr@{\,$\pm$\,}ll}
\toprule
\multicolumn{1}{c}{Quantity} & 
\multicolumn{2}{c}{Value} & 
\multicolumn{1}{c}{Units} \\
\midrule
$\nu_{\rm orb}$ & $0.003940$ & $0.000016$ & d$^{-1}$ \\
$(a_1\sin i)/c$ & $204.8$ & $25.8$ & s \\
$e$ & $0.67$ & $0.04$ & \\
$\phi_{\rm p}$ & {$0.80$} & {$0.02$} & \\
$\varpi$ & $5.28$ & $0.05$ & rad \\
$f(m_1,m_2,\sin i)$ & $0.14308$ & $0.05410$ & M$_\odot$ \\
\bottomrule
\end{tabular}
\label{tab:07}
\end{table}

\section{Discussion} 
\label{sec:5}

In this work, we have primarily focussed on deriving full orbital solutions for highly eccentric binaries. Our first example, KIC\,9651065, was also studied in our previous work \citep{murphyetal2014}, and so a direct evaluation of the improvement in technique is possible.

\subsection{Improvement in technique} 

The improvement can be seen in two ways. Firstly, the quality of the fit of the theoretical time delay curve to the data can be evaluated in terms of the reduced $\chi^2$ parameter. The former method gave a value of 2.21 for KIC\,9651065, compared to 1.80 for the analytical approach presented in this work. Secondly, one can compare the uncertainties in the orbital parameters obtained by each method. Table\,\ref{tab:08} shows that smaller uncertainties, particularly in $\varpi$, result from fitting the time delays directly, rather than fitting the radial velocities obtained by pairwise differences of the time delay data.

There are also new improvements in the elimination of systematic errors. Previously, the eccentricity would be underestimated due to the reliance on the approximation in equation (\ref{eq:13}) (equation 5 in \citealt{murphyetal2014}). This was also strongly subject to noise spikes in the Fourier transform of the time delays. Now, that approximation is only used as an initial guess, and the search for the minimum in the $\chi^2$ distribution obtains the best-fitting value more reliably.

\begin{table*} 
\centering
\caption[]{Comparison of the uncertainties in the orbital parameters for KIC\,9651065: (i) Those calculated here by fitting the time delay data in this work, vs. (ii) those calculated through fitting radial velocities obtained by taking pairwise differences of the time delay data in previous work.}
\begin{tabular}{llr@{\,$\pm$\,}lr@{\,$\pm$\,}l}
\toprule
\multicolumn{1}{l}{Quantity} & 
\multicolumn{1}{l}{Units} &
\multicolumn{2}{c}{Value} & 
\multicolumn{2}{c}{Value} \\
\multicolumn{2}{c}{\phantom{blah}} &
\multicolumn{2}{c}{This work} &
\multicolumn{2}{c}{Previous work}\\
\midrule
$\nu_{\rm orb}$ & d$^{-1}$ & $0.003684$ & $0.000011$ & $0.003667$ & $0.000016$ \\
$(a_1\sin i)/c$ & s & $183.2$ & $5.0$ & $185.0$ & $10.0$ \\
$e$ & & $0.44$ & $0.02$ & $0.47$ & $0.03$ \\
$\varpi$ & rad & $2.11$ & $0.05$ & $2.01$ & $0.30$ \\
$f(m_1,m_2,\sin i)$ & M$_\odot$ & $0.0896$ & $0.0074$ & $0.0916$ & $0.0108$ \\
\multicolumn{1}{l}{$\chi^2/N$} & & \multicolumn{2}{c}{1.80} & \multicolumn{2}{c}{2.21} \\
\bottomrule
\end{tabular}
\label{tab:08}
\end{table*}

\subsection{Factors affecting the minimum measurable time delay} 

The detection of the smallest companions, which give rise to the smallest time delays, requires a thorough understanding of the dominant contributors to the noise and how that noise can be mitigated.

There are many ways that the noise level is affected by the properties of the pulsation and/or the sampling. The cadence of the observations has little impact on the quoted 20-s limit because \textit{Kepler} observations were mostly made in a single cadence (long-cadence at 30\,min), though for stars with ample short-cadence (60-s) data the phase errors can be reduced by a factor $\sim5$ \citep{murphy2012}. The noise can be reduced when the star oscillates in many modes, providing they have similar amplitudes to the highest amplitude mode. The noise in the weighted average time delays is then reduced, though taking the weighted average means that the inclusion of more modes of much lower amplitudes than the strongest mode does not help, since phase uncertainties scale inversely with amplitude. Also for this reason, we do not consider modes with amplitudes below one tenth of that of the strongest mode in each star, and high-amplitude pulsators are clearly more favourable. Furthermore, we consider a maximum of nine modes per star due to diminishing return in computation time. Finally, it is noteworthy that the smallest detectable time delay variation has no theoretical dependence on the orbital period, providing the orbit is adequately sampled.

\section{Conclusions} 
\label{sec:6}

We have developed upon our previous work \citep{murphyetal2014}, where we showed how light arrival time delays can be obtained through pulsational phase modulation of binary stars. Formerly, radial velocities were calculated numerically from the time delays and the orbital parameters were obtained from the radial velocity curve. Here, we have shown how the same orbital parameters are obtainable directly from the time delays. The radial velocity curve is now provided only as a visualisation; it is not a necessary step in solving the orbit.

We will be applying this method to the hundreds of classical pulsators for which we have measured time delays with \textit{Kepler} data, with the aim of delivering a catalogue of time delay and radial velocity curves alongside orbital parameters in the near future. We likewise encourage developers and users of binary modelling codes to consider taking time delays as inputs.

\section*{Acknowledgements}
We are grateful to the entire Kepler team for such exquisite data.
We would like to thank Tim Bedding for his suggestions and encouragement to pursue a method that derives orbital parameters through direct fitting of the time delays.
We thank the anonymous referee for suggestions that improved this paper.

This research was supported by the Australian Research Council, 
and by the Japan Society for the Promotion of Science. Funding for the Stellar Astrophysics Centre is provided by the Danish National Research Foundation (grant agreement no.: DNRF106). The research is supported by the ASTERISK project (ASTERoseismic Investigations with SONG and Kepler) funded by the European Research Council (grant agreement no.: 267864).

\bibliography{PM2}

\end{document}